\documentclass[twocolumns]{IEEEtran}

\usepackage[english]{babel}
\usepackage{authblk}
\usepackage{siunitx}
\usepackage{array}
\usepackage{caption}
\usepackage{bm}
\usepackage[utf8x]{inputenc}
\usepackage[T1]{fontenc}
\usepackage[a4paper,top=3cm,bottom=2cm,left=3cm,right=3cm,marginparwidth=1.75cm]{geometry}
\usepackage{amsmath}
\usepackage{cite}
\usepackage[dvipsnames]{xcolor}
\usepackage{tabularx}
\usepackage[caption=false]{subfig} 
\usepackage{graphicx}
\usepackage{subfig}
\usepackage{multirow}
\usepackage{dblfloatfix}
\usepackage[colorinlistoftodos]{todonotes}
\usepackage[colorlinks=true, allcolors=blue]{hyperref}
\usepackage{hyperref}
\newcolumntype{P}[1]{>{\centering\arraybackslash}p{#1}}
\title{The role of direct sound spherical harmonics representation in externalization using binaural reproduction}
\author{Eran Miller}
\author{Boaz Rafaely}
\affil{Department of Electrical and Computer Engineering, Ben-Gurion University of the Negev,
Beer-Sheva 84105, Israel}
\affil{\href{eranmil@post.bgu.ac.il }{eranmil@post.bgu.ac.il }\ \ \ \href{mailto:br@bgu.ac.il} {br@bgu.ac.il}}

\begin{document}
\maketitle

\begin{abstract}
The importance of the information in the direct sound to human perception of spatial sound sources is an ongoing research topic. The classification between direct sound and diffuse or reverberant sound forms the basis of numerous studies in the field of spatial audio. In particular, parametric spatial audio representation methods use this classification and employ signal processing in order to enhance the audio quality at reproduction. However, current literature does not provide information concerning the impact of ideal direct sound representation on externalization, in the context of Ambisonics. This paper aims to assess the importance of the spatial information in the direct sound in the externalization of a sound field when using binaural reproduction. This is done in the spherical harmonics (SH) domain, where an ideal direct sound representation within an otherwise Ambisonics signal is simulated, and its perceived externalization is evaluated in a formal listening test. This investigation leads to the conclusion that externalization of a first order Ambisonics signal may be significantly improved by enhancing the direct sound component, up to a level similar to a third order Ambisonics signal.

\end{abstract}

\section{Introduction}

Spatial audio is attracting increasing attention in research and industry, for applications of virtual reality, music, telecommunication and more. Spatial audio recording and reproduction methods have been developed to deliver a 3D sound experience. This is achieved with the playback of sound via a loudspeaker array or by using a set of headphones, via binaural reproduction.\par
A popular format of spatial audio is Ambisonics \cite{gerzon1973periphony},\cite{gerzon1985ambisonics}. This is a linear non-parametric representation based on sound field decomposition into spherical harmonics (SH) of the first order. Ambisonics can encode either simulated sound fields, or measured sound fields using  recording systems such as the SoundField microphone. Using a suitable decoder, the Ambisonics signals can be played back using loudspeaker arrays or headphones. Due to the linear processing, Ambisonics does not introduce non-linear distortion, but has the limitation of low spatial resolution due to the inherent first SH order \cite{bertet2013investigation},\cite{braun2011localization}. Higher order Ambisonics (HOA) \cite{daniel2004further}, aims to achieve a more physically accurate reconstruction of the sound field, with higher spatial resolution than Ambisonics.  This, in turn, requires recording with a larger number of microphones, imposing practical constraints on the recording system. \par
With the aim of employing simple recording systems yet achieving high quality at reproduction, perceptually driven parametric spatial sound formats have emerged \cite{pulkki2007spatial,berge2010high,goodwin2008spatial}. These methods usually employ a B-format Ambisonics microphone and enhance the spatial audio quality by using signal processing that is based on the attributes of human spatial hearing. One such attribute is the classification between direct sound and diffuse sound, forming the basis for Directional Audio Coding (DirAC) \cite{pulkki2007spatial},\cite{politis2015sector}, an established parametric spatial sound representation. In DirAC, the signal  is divided into two streams, one corresponding to the directional part and the other to the diffuse part. This is done by the estimation of two main parameters for each time-frequency bin: the direction-of-arrival (DOA) and the diffuseness. In the reproduction stage, the direct sound stream is reproduced as a plane-wave arriving from the estimated DOA, while the diffuse sound stream is rendered using plane waves propagating in a wide range of directions after decorrelation. The perceptually based non-linear processing of methods such as DirAC were found to be preferable over Ambisonics \cite{vilkamo2009directional},\cite{barrett2010new} and even over HOA with limited order \cite{politis2017enhancement}, showing the potential benefit of appropriate manipulation of the direct and diffuse parts of the sound field.\par
In addition to DirAC, developed for spatial coding, other studies investigated the importance of the direct and reverberant parts of a sound field for spatial hearing. These have shown that the acoustic information in the direct sound dominates over that found in a single or in multiple reflections \cite{wallach1949precedence,litovsky1999precedence,zurek1987precedence}. Moreover, more recent studies  employing  binaural reproduction based on binaural room impulse responses (BRIRs), reveal that this dominance has a perceptual impact on a variety of attributes, such as localization, source width and timbre, for example \cite{ klockgether2013perceptual,devore2009accurate,ihlefeld2011effect}. Concerning externalization in particular, which is the focus of this work, in \cite{hassager2016role} it was shown that as long as the direct part of the BRIR is kept unchanged, the level of externalization is barely affected by spectral smoothing of the reverberant part of the BRIR. In \cite{catic2015role}, replacing the reverberant part of BRIRs with a monaural response did not affect the externalization in some cases, e.g. lateral sound sources, as long as the direct part was kept unchanged. While these experiments investigated the effect of the direct sound in a controlled manner by manipulating BRIRs, and even tested externalization, they did not investigate the importance of the direct sound in the context of spatial audio coding, and, in particular, in the context of Ambisonics. \par
Although previous studies with Ambisonics and SH representations manipulated the direct sound in the time-frequency domain (e.g. DirAC, SASC \cite{goodwin2008spatial}), they also included manipulations of the diffuse part. Therefore, conclusions regarding the explicit importance of the direct sound in the context of Ambisonics cannot be drawn from these studies. Furthermore, these studies focused mostly on the general quality and not specifically on externalization, and so conclusions regarding the explicit importance of the direct sound to Ambisonics with respect to externalization also cannot be drawn from these studies.\par
In this work, the importance for externalization of an ideal direct sound representation within an otherwise Ambisonics signal is investigated. It should be noted that while other attributes such as localization and source width are also important, this study focuses only on externalization. The study of other attributes is proposed for future work. A simulated binaural signal is manipulated using a mixed SH order scheme. This leads to a sound field that is divided into a direct component and a reverberant component. The direct part is rendered with a high SH order, leading to an ideal spatial representation, while the reverberant part is rendered with first order SH. The sound field is then rendered for headphones listening via binaural reproduction. Externalization of this mixed-order signal is then compared to a reference signal, rendered with high-order Ambisonics of order 30, and another high-order Ambisonics signal of order 3. The latter was chosen as an intermediate representation. First, to be consistent with previous studies \cite{politis2017enhancement}, second, because it represents the output of practical arrays such as the Eigenmike \cite{acoustics2013em32}, and third, because it was perceptually similar to the mixed-order signal when evaluated in a preliminary informal listening test. The hypothesis of this research was that a high-order SH representation of the direct sound will significantly enhance the externalization of an otherwise Ambisonics signal. In addition to refuting or validating this hypothesis, the aim of this research is to quantify the extent of this enhancement, and its dependence on the acoustic environment. The results indicate that an enhancement of the direct component of the sound field leads to a signal that is perceived to be more externalized than a first order Ambisonics signal, and in many cases similar to a third order Ambisonics signal for different audio content and acoustic environments.

\section{Binaural sound reproduction based on spherical harmonics  }
\label{sec2}
In this section an overview of the mathematical basis for binaural sound reproduction is presented. Consider a sound pressure function $p(k, r,\theta, \phi) $, where  $(r,\theta,\phi) $ are the standard spherical coordinates, $\theta \in [0,\pi]$ being the elevation angle, measured downwards from the Cartesian z-axis, and $\phi \in [-\pi,\pi)$ being the azimuth angle, measured counter-clockwise from the Cartesian x-axis on the xy-plane,  and $k=\omega/c$ is the wave number, with $\omega$ being the radial frequency and $c$ being the speed of sound. A representation of the pressure at a listener's left and right ears can be formulated as an integration over a sphere \cite{menzies2007nearfield}:
    \small
 \begin{equation}
   \label{1}
 p^{l , r}(k) = \int_{0}^{2\pi}\int_{0}^{\pi} a(k,\theta, \phi) h^{l,r}(k,\theta,\phi)\sin{\theta} d\theta d\phi,
 \end{equation}
 \normalsize
 where $a(k,\theta, \phi)$ is the complex amplitude density of a plane-wave of wavenumber $k$, with $(\theta, \phi) $  denoting the plane wave arrival direction.  $ h^{l,r}(k,\theta,\phi)$ are the complex amplitudes of the head related transfer functions (HRTFs), representing the frequency response at the ear canals at wavenumber $k$ due to a far-field sound source producing a plane wave arriving from direction $(\theta, \phi)$. The superscripts  $l , r$  represent the left and right ears, respectively. \par
 $ a(k,\theta, \phi)  $ and $h^{l,r}(k,\theta,\phi)$   can be  represented as a weighted sum of SH, defining their inverse spherical Fourier transform (ISFT) \cite{driscoll1994computing}:
 \small
\begin{equation} 
  \label{2}
  a(k,\theta, \phi) = \sum_{n=0}^{N_{a}} \sum_{m=-n}^{n} 	a_{nm}(k)Y_n^m(\theta,\phi),
  \end{equation}
  
  \begin{equation} 
  \label{3}
  h^{l,r}(k,\theta,\phi) = \sum_{n=0}^{N_h} \sum_{m=-n}^{n} 	h_{nm}^{l , r}(k) Y_n^m(\theta,\phi),
  \end{equation}
 \normalsize
 where $h_{nm}^{l , r}(k) $ and  $a_{nm}(k)$  are the spherical Fourier transform (SFT) coefficients and  $N_a $ and $N_h$ are the SH orders of $a$ and $h$, respectively. $a_{nm}(k)$  can be either analytically synthesized,  in the case of a virtual environment, or calculated by sampling the sound field using a spherical microphone array and, later, employing plane-wave decomposition (PWD) in the SH domain\cite{park2005sound},\cite{rafaely2004plane}. For $a_{nm}(k)$ computed from microphone array measurements, the order $N_a$ may be limited by the number of microphones. For example, for the Eigenmike array \cite{acoustics2013em32} a maximum order of $N_a=4$ can be computed.  $h_{nm}^{l , r}(k) $ can be computed from measurements or numerical simulations of  $h^{l,r}(k,\theta,\phi)$, which is typically sampled at thousands of directions \cite{evans1998analyzing},\cite{algazi2001cipic}. The typical order for HRTF databases is around $N_h=30$ \cite{bernschutz2013spherical},\cite{zhang2010insights}.  $Y_n^m (\theta,\phi)$ are a set of SH functions of order $n\geq 0$ and degree $-n \leq m \leq n,$ and   $(\cdot{})^*$ denotes the complex conjugate.\par
Substituting equations (\ref{2}) and (\ref{3}) into equation (\ref{1}),  using the approach developed by Rafaely and Avni  \cite{rafaely2010interaural}, the sound pressure function at the left and right ears can be calculated using the SH representation of the sound field and the HRTFs:	
 \begin{equation}
    \label{4}
  p^{l , r}(k) = \sum_{n=0}^{\min({N_a,N_h})} \sum_{m=-n}^{n} 	\tilde{a}_{nm}(k)^*h_{nm}^{l , r}(k), 
 \end{equation}where $\tilde{a}_{nm}(k)$ is the SFT of $  a(k,\theta, \phi)^*$, defined as:
 \begin{equation}
  \label{5}
\tilde{a}_{nm}(k) = \int_{0}^{2\pi}\int_{0}^{\pi} a(k,\theta, \phi)^* Y_n^m(\theta,\phi)\sin{\theta} d\theta d\phi.
 \end{equation} This, using the orthogonality property of the SH, ensures Parseval's relation is satisfied\cite{rafaely2015fundamentals}.\\
Ambisonics, for example, is based on spatial encoding using SH of the first order \cite{frank2015producing}, which could be obtained using a B-format microphone array. Utilizing the Ambisonics signals for binaural reproduction within equation (\ref{4}) yields the truncation of the summation at $\min({N_a,N_h})=1$. This kind of order truncation leads to the reproduction of a sound field of low spatial resolution \cite{rafaely2004plane} which could lead to undesired effects on key perceptual attributes such as:  timbral artifacts, loss of externalization and degraded localization \cite{rafaely2010interaural,bertet2007investigation,romigh2015efficient,thresh2017direct,liu2014analysis,braun2011localization} .

\section{Sound field representation using mixed SH order}

In this section the mathematical formulation for the representation of  a sound field with mixed SH order is presented. Equation (\ref{4}) can be further modified such that the direct component and the reverberant component of the sound field are represented using different orders. Assuming a source in the far field, the direct part of the sound field is composed of a single plane wave arriving from direction $(\theta_k,\phi_{k})$. In the SH domain, its amplitude density function is of the form $a_{nm}^{DIR}(k) =A(k)[Y_n^m (\theta_k,\phi_k)]^* $ \cite{rafaely2015fundamentals}. Now, the SH representation of the amplitude density function of the sound field can be written as 
\begin{equation}
    \label{6}
 {a}_{nm}(k) = a_{nm}^{DIR}(k)+a_{nm}^{REV}(k),
 \end{equation}
where $a_{nm}^{REV}(k)$ is the amplitude density function of the reverberant part of the sound field. This leads to the mixed SH formulation of the sound field:

\begin{equation}
\begin{split}
    \label{7}
  p^{l , r}(k) = \sum_{n=0}^{N_{d}} \sum_{m=-n}^{n} 	[\tilde{a}_{nm}^{DIR}(k)]^*h_{nm}^{l , r}(k) \\
  +\sum_{n=0}^{N_{r}} \sum_{m=-n}^{n}[\tilde{a}_{nm}^{REV}(k)]^*h_{nm}^{l , r}(k),
  \end{split}
 \end{equation}
 where $N_d$  and $N_r$ are the orders for the direct and reverberant components, respectively, and are not necessarily equal, allowing for the reproduction of spatial audio with enhanced direct sound.

\begin{table*}[ht]
\centering
\begin{tabular}{|l|l|l|}
\hline
 &\textbf{Environment 1} &\textbf{Environment 2 } \\ \hline
 \textbf{Signal type}&Noise , Speech & Noise , Speech\\ \hline
 \textbf{DRR}&$-3.52\,$dB  & $-9.52\,$dB \\ \hline
 \textbf{Source-listener dist.}&$3.315\,$m  &$6.63\,$m  \\ \hline
\end{tabular}
 \caption{Parameters of the acoustic environments}
 \label{table: table1}
\end{table*}

\begin{table*}
\centering
\begin{tabular}{|l|l|l|}
\hline
 \textbf{Signal}&\bm{$N_d$} &\bm{$N_r$} \\ \hline
 Mixed order&30  &1  \\ \hline
 Reference&30  &30  \\ \hline
 Third order&3  &3  \\ \hline
 Anchor&1  &1  \\ \hline
\end{tabular}
 \caption{Binaural signals SH order}
 \label{table: table2}
\end{table*}

\section{Methodology}

With the aim of studying the importance for externalization of the information in the direct sound using binaural reproduction, a listening test based on Recommendation ITU-R BS.1534-1 (MUSHRA, MUltiple Stimuli with Hidden Reference and Anchor) was developed. \par
A rectangular room of dimensions  15.5 $\times$ 9.8 $\times$ 7.5 m with a wall reflection coefficient of $R = 0.8$  and 	$T_{60} = 0.75\,$s was simulated using the image method\cite{allen1979image}. The critical distance for this room is $r_d = 2.21 m$.  A room impulse response from a point source to a listener's position was calculated and represented in the form of $a(t,\theta,\phi)$, and further encoded in the SH domain as $a_{nm}(t)$, with $t$ representing time. The sound field,  $a_{nm}(k)$,  was then calculated at the listener's position with orders $N_d$ and $N_r$ according to equation (\ref{7}), by convolving the room impulse response $a_{nm}(t)$ with the source signal, $s(t)$, leading to $a_{nm}(k)$ after transformation to frequency. $a_{nm}(k)$ was later multiplied with a set of HRTFs of matching orders $N_d$ and $N_r$ in order to provide a binaural signal. For this investigation, the  Cologne HRTF compilation of the Neumann KU-100 \cite{bernschutz2013spherical} was used.\par
 In order to asses the impact of the direct part to the perceived externalization, four binaural signals were generated:\par
 
\begin{enumerate}
\item \textbf{Mixed order signal}: a binaural signal with an ideal representation of the direct sound, encoded with $N_d=30$,  and low order representation of the reverberant sound encoded with  $N_r = 1$. 
\item \textbf{Reference}: a binaural signal with an ideal representation of both the direct and reverberant sound components. This was implemented in practice using encoding with a SH order of $N_d = N_r = 30$ . This high SH order represents the most accurate representation of the binaural signal for this system. It also avoids spatial aliasing in the range of $f \leq 20$ kHz \cite{rafaely2015fundamentals} and also avoids the need for other methods of interpolation for producing HRTFs directions that are unavailable in the database. Furthermore, according to \cite{ahrens2012hrtf}, HRTF representation of such a high SH order should be sufficient to yield correct spatial details.
\item \textbf{Anchor}:  a signal encoded with $N_d = N_r =1$, representing  encoding  in the Ambisonics format.
\item \textbf{Third order signal} $N_d=Nr=3$  was chosen as an intermediate representation, providing an additional reference point.\\
\end{enumerate}
\par
The binaural signals were generated in two virtual acoustic environments, referred to as \textbf{environment 1} and \textbf{environment 2}. The two different environments were chosen to diversify the acoustic environment condition, to ensure that  conclusions are not specific to a single environment. The two environments differ in the source  position relative to the listener, which was located at position [9,7,1.7].  For environment 1, the source was located at 1.5 times the critical distance from the listener's location (i.e $3.315\,$m), while for environment 2, the source was located at 3 times the critical distance from the listener's location (i.e. $6.63\,$m), leading to a reduction of $6\,$dB in the direct sound energy relative to the reverberant sound energy,  compared to environment 1. It should be noted that both environments represent relatively reverberant conditions, as the listener is positioned further away from the source compared to the  critical distance. This setting was chosen due to its improved externalization, and for studying the effect of enhancing the direct sound in acoustic environments with distant sources and  negative DRR.  For both environments, the source was located at $30^\circ$  from the listener, relative to the HRTF coordinate system, and at the same height as the listener's head. An important goal of this experiment is to evaluate the externalization of the mixed order signal in both environments, which differ in their direct-to-reverberant ratio (DRR).\par
Two source signals were used:
\begin{enumerate}
\item A pink noise repeating burst (1s duration including 20ms fade in and fade out, followed by a 0.3s pause before the next burst), chosen for its  wide bandwidth. 
\item Speech segment in the English language (3.26s duration) from the TIMIT corpus \cite{garofolo1993darpa}, chosen as it represents a typical real life audio content.\par 
\end{enumerate}
 The different environments and signals are summarized in tables \ref{table: table1} and \ref{table: table2}, respectively.\par
Previous studies \cite{avni2013spatial} have shown that the truncation of the SH series to a lower order may alter the timbre of a binaural signal. This may affect the task of rating the signals according to externalization level only. In order to overcome this issue, a spectral equalization filter was employed, as described in \cite{ben2017spectral}, ensuring all signals were equalized to the reference signal. The signals were rendered using the Sound Scape Renderer (SSR) software \cite{geier2008soundscape} and played back using AKG701 reference headphones. All signals were convolved with a matching headphone compensation filter. For spatial realism, horizontal head movements were allowed, and the headphones were mounted with a Razor AHRS head tracker. The signals were generated to support head rotations, covering the horizontal plane with a resolution of $1^\circ{}$. The latency for the SSR, under the settings used in this experiment, is  $\sim 17.5\,$ms. Together with the latency of the head tracker, the total latency is lower than 60ms, which is sufficient for acceptable levels of localization accuracy \cite{brungart2004interaction}. \par
15 normal hearing subjects participated in this experiment. 6 of them are expert listeners and the rest are naive listeners.  The experiment included a total of 4 MUSHRA screens:  2 screens for each of the two environments, according to table (\ref{table: table1}). 
Each screen presented 4 signals according to table (\ref{table: table2}). In each screen the subjects were instructed to rate the degree to which the sound source is perceived to be originating from inside or outside the head, compared to the reference, on a scale from 0-100. Before rating, the subjects conducted a training task, in order to familiarize them with the experiment and stimuli. 

 \begin{figure*}[hpt]
\begin{center}
\label{fig_main}
\centering
  \subfloat[Environment 1 : Noise signal.]{\includegraphics[width=0.5\textwidth]{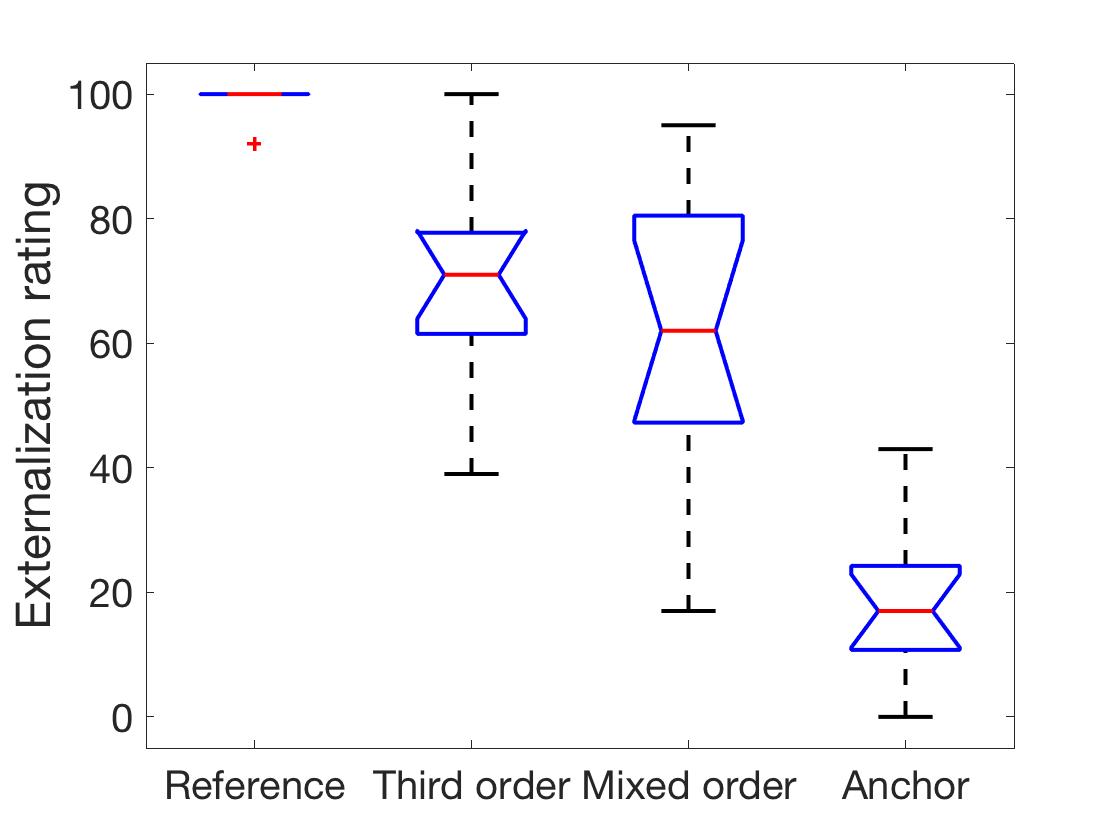}\label{fig1:f1}}
    \subfloat[Environment 1 : Speech signal.]{\includegraphics[width=0.5\textwidth]{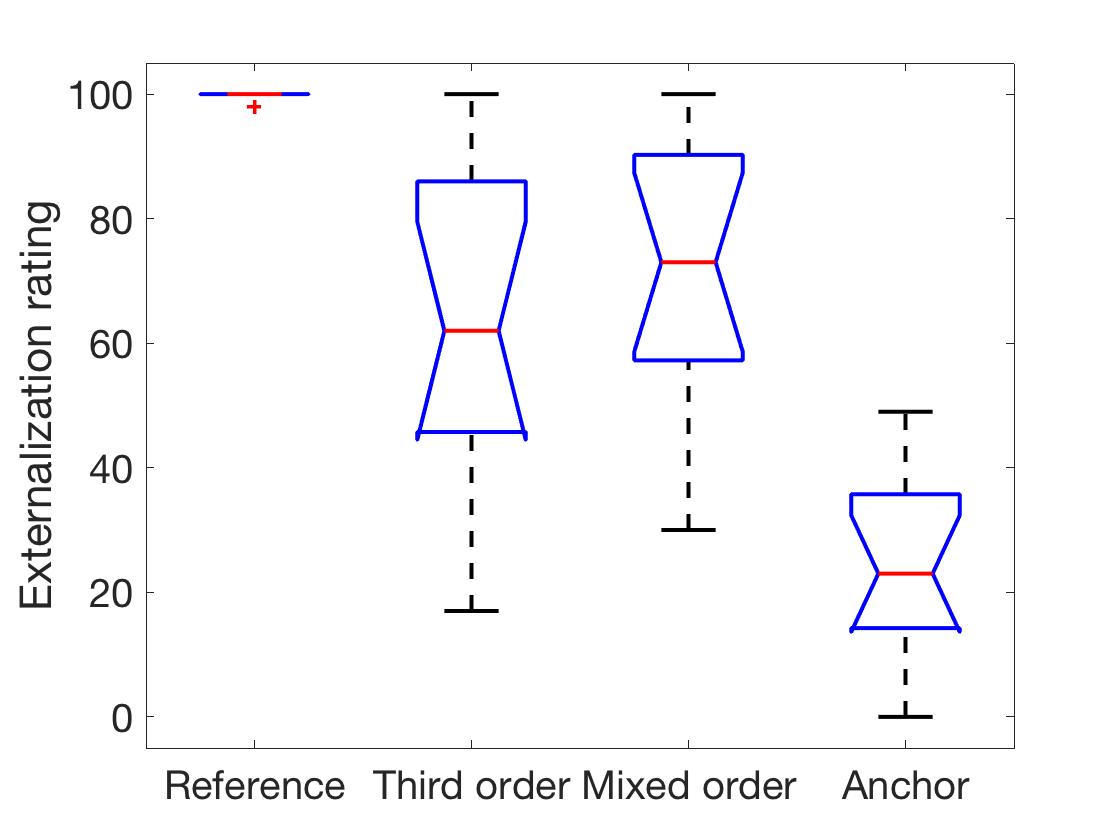}\label{fig1:f2}}
      \hfill
   \subfloat[Environment 2 : Noise signal.]{\includegraphics[width=0.5\textwidth]{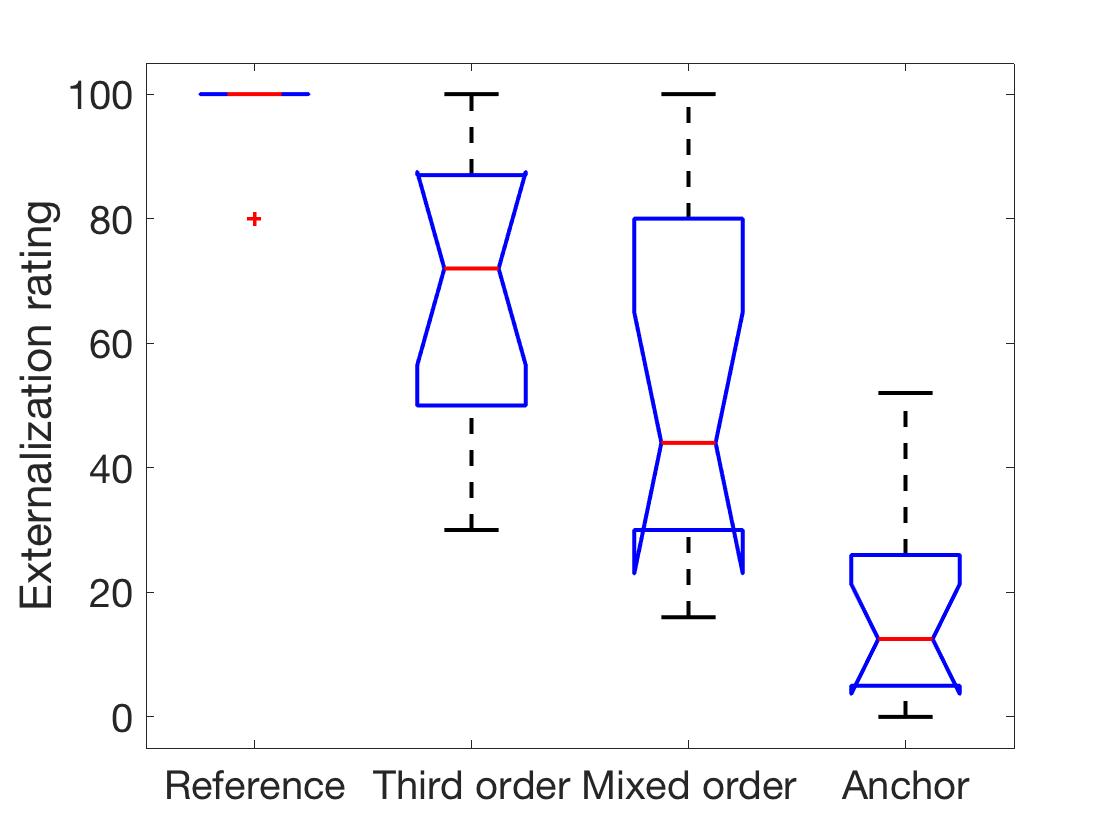}\label{fig1:f3}}
    \subfloat[Environment 2 : Speech signal.]{\includegraphics[width=0.5\textwidth]{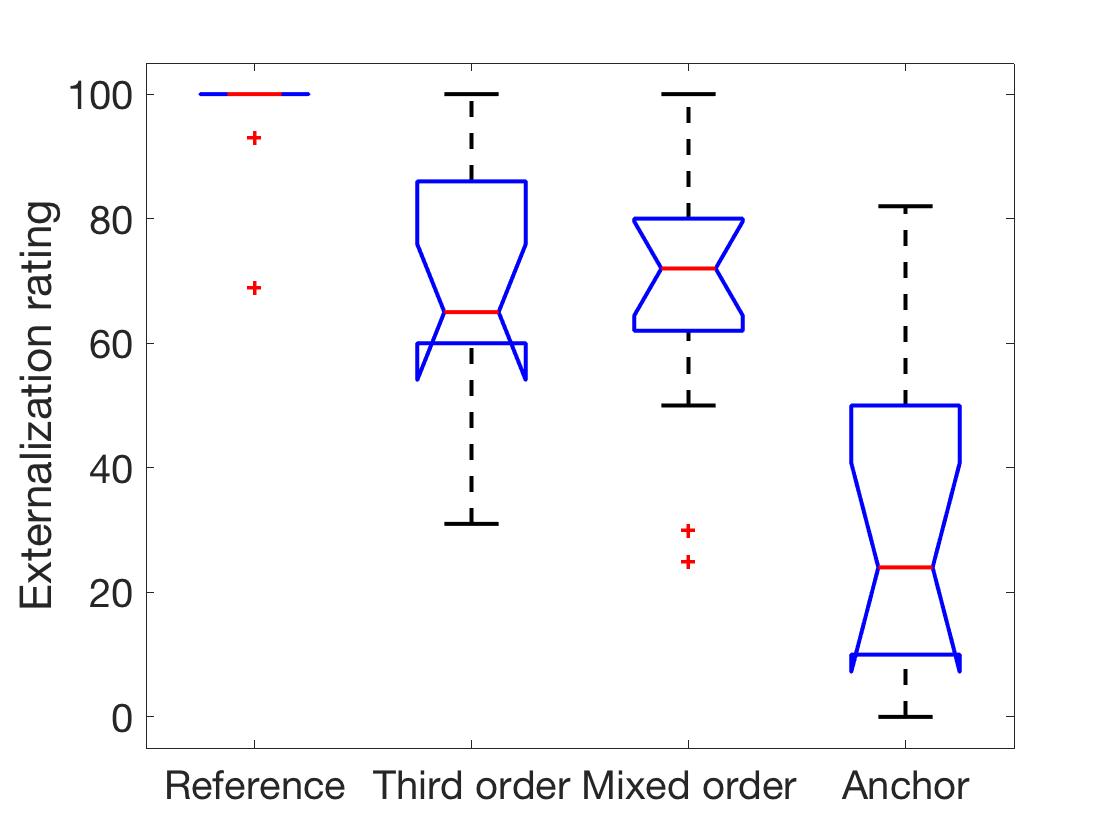}\label{fig1:f4}}
    \hfill
   \captionsetup{justification=centering}
  \caption{Results for the externalization ratings in environment 1 and environment 2. Box plot visualization, marking the median with a red line where the bottom and top edges of the box represent the 25th and 75th percentiles, respectively. The whiskers represent the variability of the ratings outside the upper and lower quartiles. Outliers are marked with red '+'.  The width of the box plot notches has been calculated such that boxes with non-overlapping notches have medians which are different at the 5$\%$ significance level}
  \end{center}
\end{figure*}

\section{Results}

Externalization results are presented in this section, divided into the two environments. The one-way analysis of variance (ANOVA) was used as the statistical test. In order to \\

\subsubsection{Environment 1}

Figures \ref{fig1:f1} and \ref{fig1:f2} present the results for environment 1.  The reference is shown to be perceived as the most externalized for both source signals. The median scores of all signals differ with significance, with p < 0.001  for both speech and noise, except for the difference of the scores between the mixed order signal and the third order signal, which is not statistically significant with p = 0.33 for the noise source and p = 0.36 for the speech source.

\subsubsection{Environment 2} 

Figures \ref{fig1:f3} and  \ref{fig1:f4} present the results for environment 2. The reference is shown to be perceived as the most externalized. The median scores of all signals differ with significance (p < 0.001  for both speech and noise), except for the mixed order signal and the third order signal.  As in environment 1, no statistically significant difference was found between the  scores of the mixed order signal and  the third order signal. The p-value between these signals are  p  = 0.11 for the noise source and  p = 0.89 for the speech source.

\section{Discussion }

The effect of enhancing the spatial representation of the direct part of a sound field is shown to impact the externalization of a binaural signal, as the mixed order signal was perceived to be more externalized than the first order signal in both environments and for both noise and speech. This proves that even with a decrease in the DRR (see table \ref{table: table1}), the information encoded in the direct part contributes to externalization.\par 
The scores for the mixed order signal and the third order signal were found to be of no statistically significant difference. This shows that to some extent, with respect to externalization, enhancing the direct part in a first-order Ambisonics signal could yield a signal which resembles a third order Ambisonics signal. This is in some agreement with the results presented in \cite{politis2017enhancement}, comparing DirAc to Ambisonics for overall quality. \par    
Because the externalization of the reference signal was not compared directly between the two environments, it is not possible to deduce which of the two environments is perceived to be more externalized. Nevertheless, the effect of spherical harmonics order (for the four signal types) seems similar under the two environments. Indeed, a multi-factor ANOVA test confirms that the interaction between the spherical harmonics order factor and the environment factor is not significant. These results can be explained by the similar nature of the two environments. They are both  reverberant, with the listener's distance from the source being greater than the critical.\par
 It is interesting to note that the score for the mixed order signal for both environments, seems to slightly improve compared to the third order, when comparing speech to noise. In environment 2 (see Figs. \ref{fig1:f3},\ref{fig1:f4}), for example, for the noise source, the mixed order median score is 44 while the third order signal score is 72. On the other hand, for the speech source, the score of the mixed order signal is 72 while the score of the third order signal is 65. A possible explanation could be the non-stationarity of speech, which includes repeated onsets, compared to the noise signal that has a single onset every 1s. Therefore, with speech (or any other signal that contains numerous onsets), the direct sound may have a greater importance and its enhancement may therefore lead to a greater impact on externalization.

\section{Conclusion  }

In this paper, a binaural signal was manipulated in the SH domain, facilitating an evaluation of the improvement in the externalization of a sound field that is composed of a reverberant part of low spatial resolution and a direct part of ideal spatial representation. A subjective listening test showed that an enhancement of the direct part alone yields a sound field which is perceived to be externalized at a level similar to that of a sound field represented in the third order, and more externalized than a sound field represented in the first order for all cases inspected. The results presented in this paper could have implications for various applications of spatial audio:\\
\begin{enumerate}
\item For applications in which the room impulse response is measured with a first-order SH microphone array such as the Soundfield microphone \cite{farrar1979soundfield}, the benefit to externalization by rendering the direct sound with a high spatial quality is shown to be significant as quantified in this paper. Room acoustics auralization and binaural playback of anechoic audio signals through measured room impulse responses could be example applications. \\
\item For applications in which audio signals are directly recorded with a microphone array of a first order, enhancing the spatial resolution of the direct sound may require a preliminary stage of identifying direct-sound components. Although this approach is already implemented in DirAC \cite{pulkki2007spatial}, new methods that were recently developed for identifying direct-sound components in the time-frequency domain, can be studied to further improve the quality of Ambisonics signals, motivated by the potential benefit to externalization, as presented here. These methods include the direct-path-dominance (DPD) test \cite{nadiri2014localization}, with its recent extensions \cite{rafaely2018speaker}\cite{schymura2018exploiting}. \\
\end{enumerate}
Similar studies, concerning other attributes, such as  localization, source width, timbre perception, and more, are proposed for future work.

\bibliographystyle{unsrt}
\bibliography{Biblio1}

\end{document}